


\font\twelverm=cmr10  scaled 1200   \font\twelvei=cmmi10  scaled 1200
\font\twelvesy=cmsy10 scaled 1200   \font\twelveex=cmex10 scaled 1200
\font\twelvebf=cmbx10 scaled 1200   \font\twelvesl=cmsl10 scaled 1200
\font\twelvett=cmtt10 scaled 1200   \font\twelveit=cmti10 scaled 1200
\font\twelvesc=cmcsc10 scaled 1200
\skewchar\twelvei='177   \skewchar\twelvesy='60


\def\twelvepoint{\normalbaselineskip=12.4pt plus 0.1pt minus 0.1pt
  \abovedisplayskip 12.4pt plus 3pt minus 9pt
  \belowdisplayskip 12.4pt plus 3pt minus 9pt
  \abovedisplayshortskip 0pt plus 3pt
  \belowdisplayshortskip 7.2pt plus 3pt minus 4pt
  \smallskipamount=3.6pt plus1.2pt minus1.2pt
  \medskipamount=7.2pt plus2.4pt minus2.4pt
  \bigskipamount=14.4pt plus4.8pt minus4.8pt
  \def\rm{\fam0\twelverm}          \def\it{\fam\itfam\twelveit}%
  \def\sl{\fam\slfam\twelvesl}     \def\bf{\fam\bffam\twelvebf}%
  \def\mit{\fam 1}                 \def\cal{\fam 2}%
  \def\sc{\twelvesc}               \def\tt{\twelvett}
  \def\sf{\twelvesf}
  \textfont0=\twelverm   \scriptfont0=\tenrm   \scriptscriptfont0=\sevenrm
  \textfont1=\twelvei    \scriptfont1=\teni    \scriptscriptfont1=\seveni
  \textfont2=\twelvesy   \scriptfont2=\tensy   \scriptscriptfont2=\sevensy
  \textfont3=\twelveex   \scriptfont3=\twelveex  \scriptscriptfont3=\twelveex
  \textfont\itfam=\twelveit
  \textfont\slfam=\twelvesl
  \textfont\bffam=\twelvebf \scriptfont\bffam=\tenbf
  \scriptscriptfont\bffam=\sevenbf
  \normalbaselines\rm}



\def\beginlinemode{\endmode
  \begingroup\parskip=0pt \obeylines\def\\{\par}\def\endmode{\par\endgroup}}
\def\beginparmode{\endmode
  \begingroup \def\endmode{\par\endgroup}}
\let\endmode=\par
{\obeylines\gdef\
{}}
\def\singlespace{\baselineskip=\normalbaselineskip}

\def\oneandahalfspace{\baselineskip=\normalbaselineskip
  \multiply\baselineskip by 3 \divide\baselineskip by 2}
\def\doublespace{\baselineskip=\normalbaselineskip \multiply\baselineskip by 2}

\newcount\firstpageno
\firstpageno=2
\footline={\ifnum\pageno<\firstpageno{\hfil}\else{\hfil\twelverm\folio\hfil}\fi}

\def\toppageno{\global\footline={\hfil}\global\headline
  ={\ifnum\pageno<\firstpageno{\hfil}\else{\hfil\twelverm\folio\hfil}\fi}}
\let\rawfootnote=\footnote              
\def\footnote#1#2{{\rm\singlespace\parindent=0pt\parskip=0pt
  \rawfootnote{#1}{#2\hfill\vrule height 0pt depth 6pt width 0pt}}}
\def\raggedcenter{\leftskip=4em plus 12em \rightskip=\leftskip
  \parindent=0pt \parfillskip=0pt \spaceskip=.3333em \xspaceskip=.5em
  \pretolerance=9999 \tolerance=9999
  \hyphenpenalty=9999 \exhyphenpenalty=9999 }
\def\dateline{\rightline{\ifcase\month\or
  January\or February\or March\or April\or May\or June\or
  July\or August\or September\or October\or November\or December\fi
  \space\number\year}}
\def\received{\vskip 3pt plus 0.2fill
 \centerline{\sl (Received\space\ifcase\month\or
  January\or February\or March\or April\or May\or June\or
  July\or August\or September\or October\or November\or December\fi
  \qquad, \number\year)}}


\hsize=6.5truein
\hoffset=0.0truein
\vsize=8.5truein
\voffset=0.25truein
\parskip=\medskipamount
\toppageno
\twelvepoint
\doublespace
\def\\{\cr}
\overfullrule=0pt 



\def\tutp#1{
  \rightline{\rm TUTP--#1}} 

\def\title#1{                   
   \null \vskip 3pt plus 0.3fill \beginlinemode
   \doublespace \raggedcenter {\bf #1} \vskip 3pt plus 0.1 fill}

\def\author                     
  {\vskip 3pt plus 0.1fill \beginlinemode \doublespace \raggedcenter}

\def\affil                      
  {\vskip 3pt \beginlinemode \doublespace \raggedcenter \it}

\def\abstract                   
  {\vskip 3pt plus 0.1fill \subhead {Abstract:}
   \beginparmode \narrower \oneandahalfspace }

\def\endtopmatter               
  {\vskip 3pt plus 0.1fill \endpage \body}

\def\body                       
  {\beginparmode}               

\def\head#1{                    
   \goodbreak \vskip 0.4truein  
  {\immediate\write16{#1} \raggedcenter {\sc #1} \par}
   \nobreak \vskip 3pt \nobreak}

\def\subhead#1{                 
  \vskip 0.25truein             
  {\raggedcenter {\it #1} \par} \nobreak \vskip 3pt \nobreak}

\def\beneathrel#1\under#2{\mathrel{\mathop{#2}\limits_{#1}}}

\def\refto#1{${\,}^{#1}$}       

\newdimen\refskip \refskip=0pt
\def\references         
  {\head{References}    
   \beginparmode \frenchspacing \parindent=0pt \leftskip=\refskip
   \parskip=0pt \everypar{\hangindent=20pt\hangafter=1}}

\gdef\refis#1{\item{#1.\ }}                     

\gdef\journal#1, #2, #3 {               
    {\it #1}, {\bf #2}, #3.}            




\def\endreferences{\body}

\def\figurecaptions             
  {\endpage \beginparmode \head{Figure Captions}
   \parskip=3pt \everypar{\hangindent=20pt\hangafter=1} }

\def\endpage                    
  {\vfill\eject}

\def\endpaper   {\endmode\vfill\supereject}
\def\endjnl     {\endpaper\end}


\def\ref#1{ref.{#1}}                    
\def\Ref#1{Ref.{#1}}                    
\def\[#1]{[\cite{#1}]}
\def\cite#1{{#1}}


\def\(#1){(\call{#1})}
\def\call#1{{#1}}
\def\frac#1#2{{#1 \over #2}}
\def\half{  {\frac 12}}

\def\12{{1\over2}}

\def\sla{\raise.15ex\hbox{$/$}\kern-.57em}
\def\leaderfill{\leaders\hbox to 1em{\hss.\hss}\hfill}
\def\twiddle{\lower.9ex\rlap{$\kern-.1em\scriptstyle\sim$}}
\def\bigtwiddle{\lower1.ex\rlap{$\sim$}}
\def\gtwid{\mathrel{\raise.3ex\hbox{$>$\kern-.75em\lower1ex\hbox{$\sim$}}}}
\def\ltwid{\mathrel{\raise.3ex\hbox{$<$\kern-.75em\lower1ex\hbox{$\sim$}}}}
\def\square{\kern1pt\vbox{\hrule height 1.2pt\hbox{\vrule width 1.2pt\hskip 3pt
   \vbox{\vskip 6pt}\hskip 3pt\vrule width 0.6pt}\hrule height 0.6pt}\kern1pt}
\def\tdot#1{\mathord{\mathop{#1}\limits^{\kern2pt\ldots}}}

\def\pmb#1{\setbox0=\hbox{#1}%
  \kern-.025em\copy0\kern-\wd0
  \kern  .05em\copy0\kern-\wd0
  \kern-.025em\raise.0433em\box0 }

\catcode`@=11
\newcount\r@fcount \r@fcount=0
\newcount\r@fcurr
\immediate\newwrite\reffile
\newif\ifr@ffile\r@ffilefalse
\def\w@rnwrite#1{\ifr@ffile\immediate\write\reffile{#1}\fi\message{#1}}

\def\writer@f#1>>{}
\def\referencefile{
  \r@ffiletrue\immediate\openout\reffile=\jobname.ref%
  \def\writer@f##1>>{\ifr@ffile\immediate\write\reffile%
    {\noexpand\refis{##1} = \csname r@fnum##1\endcsname = %
     \expandafter\expandafter\expandafter\strip@t\expandafter%
     \meaning\csname r@ftext\csname r@fnum##1\endcsname\endcsname}\fi}%
  \def\strip@t##1>>{}}

\def\citeall#1{\xdef#1##1{#1{\noexpand\cite{##1}}}}
\def\cite#1{\each@rg\citer@nge{#1}}     

\def\each@rg#1#2{{\let\thecsname=#1\expandafter\first@rg#2,\end,}}
\def\first@rg#1,{\thecsname{#1}\apply@rg}       
\def\apply@rg#1,{\ifx\end#1\let\next=\relax
\else,\thecsname{#1}\let\next=\apply@rg\fi\next}

\def\citer@nge#1{\citedor@nge#1-\end-}  
\def\citer@ngeat#1\end-{#1}
\def\citedor@nge#1-#2-{\ifx\end#2\r@featspace#1 
  \else\citel@@p{#1}{#2}\citer@ngeat\fi}        
\def\citel@@p#1#2{\ifnum#1>#2{\errmessage{Reference range #1-#2\space is bad.}
    \errhelp{If you cite a series of references by the notation M-N, then M and
    N must be integers, and N must be greater than or equal to M.}}\else%
 {\count0=#1\count1=#2\advance\count1
by1\relax\expandafter\r@fcite\the\count0,%
  \loop\advance\count0 by1\relax
    \ifnum\count0<\count1,\expandafter\r@fcite\the\count0,%
  \repeat}\fi}

\def\r@featspace#1#2 {\r@fcite#1#2,}    
\def\r@fcite#1,{\ifuncit@d{#1}          
    \expandafter\gdef\csname r@ftext\number\r@fcount\endcsname%
    {\message{Reference #1 to be supplied.}\writer@f#1>>#1 to be supplied.\par
     }\fi%
  \csname r@fnum#1\endcsname}

\def\ifuncit@d#1{\expandafter\ifx\csname r@fnum#1\endcsname\relax%
\global\advance\r@fcount by1%
\expandafter\xdef\csname r@fnum#1\endcsname{\number\r@fcount}}

\let\r@fis=\refis                       
\def\refis#1#2#3\par{\ifuncit@d{#1}
    \w@rnwrite{Reference #1=\number\r@fcount\space is not cited up to now.}\fi%
  \expandafter\gdef\csname r@ftext\csname r@fnum#1\endcsname\endcsname%
  {\writer@f#1>>#2#3\par}}

\def\r@ferr{\endreferences\errmessage{I was expecting to see
\noexpand\endreferences before now;  I have inserted it here.}}
\let\r@ferences=\references
\def\references{\r@ferences\def\endmode{\r@ferr\par\endgroup}}

\let\endr@ferences=\endreferences
\def\endreferences{\r@fcurr=0
  {\loop\ifnum\r@fcurr<\r@fcount
    \advance\r@fcurr by 1\relax\expandafter\r@fis\expandafter{\number\r@fcurr}%
    \csname r@ftext\number\r@fcurr\endcsname%
  \repeat}\gdef\r@ferr{}\endr@ferences}


\let\r@fend=\endpaper\gdef\endpaper{\ifr@ffile
\immediate\write16{Cross References written on []\jobname.REF.}\fi\r@fend}

\catcode`@=12

\citeall\refto          
\citeall\ref            %
\citeall\Ref            %

\doublespace
\vglue 0. truein
\tutp{94-10}

\title
{
Electroweak Dyons
}
\smallskip
\author
{Tanmay Vachaspati}
\affil
{
Tufts Institute of Cosmology, Department of Physics and Astronomy,
Tufts University, Medford, MA 02155.
}

\abstract
\doublespace

We consider dyon configurations in the standard electroweak model.
In the presence of a $\theta$ term and no fermions, the usual
result $q = (n+\theta /2\pi ) e$ is obtained for the electric charge
spectrum. The effect of including standard model fermions
is discussed qualitatively.

\endtopmatter

\noindent{\bf 1. Introduction}

It is often not appreciated that the standard model of the electroweak
interactions contains magnetic monopoles\refto{yn}. The only way
in which an electroweak monopole is different from a usual monopole is
that it is always connected by a string to an antimonopole. But this string
is {\it  not} a Dirac or other string that surreptitiously returns the
magnetic flux; instead it is an electromagnetically neutral string made up
of $Z$ magnetic field\refto{tv1}. And, as originally asserted by Nambu,
electroweak monopoles are genuine magnetic monopoles for which the divergence
of $\vec B$ does not vanish.

At the same time, the electroweak monopole is mysterious in some ways.
For example, at first sight, the Dirac quantization condition seems to be
violated. This is because the magnetic charge on the electroweak monopole is:
$$
m = {{4\pi} \over e} sin^2 \theta_W
\eqno (1.1)
$$
where, $\theta_W$ is the Weinberg angle and $e$ the electric
charge of the electron.
In contrast to the Dirac quantization condition, $e m \ne 4\pi n$ for any
integer $n$ for general $\theta_W$.

In addition, if one generalizes the electroweak monopole to electroweak
dyons, a naive analysis leads one to expect fractional electric
charge on the dyon even in the absence of CP violation. To see this,
one uses the quantization condition relevant for two dyons with magnetic
charges $m_1$, $m_2$ and electric charges $q_1$, $q_2$:
$$
q_1 m_2 - q_2 m_1 = 4\pi n \ .
\eqno (1.2)
$$
If we consider a monopole and an antimonopole, such as would be present
at the ends of a string, together with eqn. (1.1), we get,
$$
q_1 + q_2 = {{ne} \over {sin^2 \theta_W}} \ .
\eqno (1.3)
$$
Disregarding the fermions in the theory, CP is conserved, and dyons
with charge $(m_1, - q_1)$ also exist.
Then (1.3) together with the $q_1 \rightarrow - q_1$ version of (1.3),
gives the electric charge spectrum of the dyon:
$$
q = n {{e} \over {sin^2 \theta_W}} \ ,
\ \ \ \left ( n + \half \right ) {{e} \over {sin^2 \theta_W}}.
\eqno (1.4)
$$
(If we assume that $q=0$ is in the spectrum, the half integer solutions
will be eliminated.) Hence it appears as though the dyon spectrum
is not quantized in units of $e$.

As first described by Witten\refto{ew},
on including a $\theta$ term in the bosonic sector of a theory with
magnetic monopoles, the monopoles get an electric charge $e\theta / 2\pi$.
Since electroweak monopoles are
genuine magnetic monopoles, we expect the same phenomenon to occur and
electroweak monopoles to get Witten's value of the electric charge.
However, if we were to simply add $e\theta /2\pi$ to the charge that
appears in (1.4), the periodicity under $\theta \rightarrow \theta + 2\pi$
would be lost. Clearly, something is amiss.

In this paper, we would like to address some of these issues. To start
with we outline the electroweak monopole system. In Sec. 2, one of
our main goals is to emphasize Nambu's assertion that electroweak
monopoles should be thought of as genuine monopoles. We then turn to
the construction of dyons at a classical level in Sec. 3.
For this we exhibit
field configurations that satisfy the asymptotic field equations and
have dyonic field strengths. At the classical level, the dyon can have
any electric charge but, by quantizing the angular momentum,
we expect to get a discrete dyon charge spectrum. This mission is
accomplished in Sec. 4 where we find
that the electric charge on the dyon is quantized in units of $e$ as
one would have hoped for and not as eqn. (1.4) would have us believe.
We also include a $\theta$ term in the action (without fermions)
and find that electroweak dyons carry the Witten value of electric
charge exactly as
ordinary monopoles do. In Sec. 5 we discuss what might happen to
the dyon charge spectrum due to CP violation in the standard model.
Finally we explain the significance that the Witten electric charge on an
electroweak monopole may have in the context of electroweak baryogenesis.

\noindent{\bf 2. Electroweak Monopoles}

In this section, we will repeat some of Nambu's arguments and derive
the magnetic charge on an electroweak monopole. For this, assume
that we have a semi-infinite string along the $-z$ axis
and a monopole at the origin.

First of all, we define the $Z$ and $A$ gauge fields
for general Higgs field $\Phi$:
$$
Z_\mu \equiv cos\theta_W ~n^a W_\mu ^a - sin\theta_W ~Y_\mu \ ,
\ \ \ \
A_\mu \equiv sin\theta_W ~n^a W_\mu ^a + cos\theta_W ~Y_\mu \ ,
\eqno (2.1)
$$
where, $W_\mu ^a$ and $Y_\mu$ denote the $SU(2)$ and $U(1)$ hypercharge
gauge fields and
$$
n^a \equiv  -{{\Phi^{\dag} \tau^a \Phi} \over {\Phi^{\dag} \Phi}} \ .
\eqno (2.2)
$$
There are several different choices for defining the electromagnetic
field strength but, following Nambu, we choose:
$$
A_{\mu \nu} = sin\theta _W ~n^a W_{\mu \nu} ^a + cos\theta _W ~Y_{\mu \nu}
\eqno (2.3)
$$
where, $W_{\mu \nu}^a$ and $Y_{\mu \nu}$ are field strengths. (The
different choices for the definition of the field strength agree in
the region where $D_\mu \Phi = 0$ where $D_\mu$ is the covariant
derivative operator.) And the combination of $SU(2)$ and
$U(1)$ field strengths orthogonal to $A_{\mu \nu}$ is defined to be the
$Z$ field strength:
$$
Z_{\mu \nu} = cos\theta _W ~n^a W_{\mu \nu} ^a - sin\theta_W ~Y_{\mu \nu} \ .
\eqno (2.4)
$$
The $A$ and $Z$ magnetic fluxes through a spatial surface will be denoted
by $F_A$ and $F_Z$ and these are given in the usual way by the surface
integrals of the field strengths. Therefore we can write
$$
F_Z = cos\theta_W F_n - sin\theta_W F_Y \ ,
\ \ \ \
F_A = sin\theta_W F_n + cos\theta_W F_Y \ ,
\eqno (2.5)
$$
where we have denoted the $SU(2)$ flux (parallel to $n^a$ in group
space) by $F_n$ and the hypercharge flux by $F_Y$.

Now consider a large sphere $\Sigma$ centered on the monopole.
The monopole-string configuration is such that there
is only an electromagnetic magnetic flux through $\Sigma$ except at
the South Pole ($S$), where there is only a $Z$ magnetic flux. Hence,
denoting the electromagnetic flux by $F_A$ and the Z flux by $F_Z$,
$$
F_Z |_\Sigma = 0 = F_A |_S   \ .
\eqno (2.6)
$$

Equation (2.6) together with (2.5) tells us that
$$
F_n |_\Sigma = tan\theta_W F_Y |_\Sigma \ , \ \ \
F_n |_S = - cot \theta_W F_Y |_S \ .
\eqno (2.7)
$$
But, the hypercharge flux must be conserved as it is a $U(1)$ flux. So
$$
F_Y |_\Sigma = - F_Y |_S \equiv F_Y \ ,
\eqno (2.8)
$$
and, inserting this and (2.7) in (2.5) gives
$$
F_A |_\Sigma = {{F_Y} \over {cos\theta_W}} \ , \ \ \
F_Z |_S = {{F_Y} \over {sin\theta_W}} \ .
\eqno (2.9)
$$

To proceed further, we need to put in some dynamics. This is most simply done
by realizing that the string along the $-z$ axis is an ordinary Nielsen-Olesen
vortex\refto{hnpo} embedded in the electroweak model\refto{tvmb}
and so the flux is quantized
in units of $4\pi /\alpha$ where $\alpha \equiv \sqrt{g^2 + g'{}^2}$ is the
coupling of the $Z$ to the Higgs field. Therefore, for the unit winding
string,
$$
F_Z |_S = {{4\pi} \over {\alpha}} \ .
\eqno (2.10)
$$
Then (2.9) yields,
$$
F_Y = {{4\pi} \over {\alpha}} sin\theta_W \ , \ \ \
F_A |_\Sigma = {{4\pi} \over {\alpha}} tan\theta_W
                  = {{4\pi} \over {e}} sin^2 \theta_W \
\eqno (2.11)
$$
where, $g = \alpha cos\theta_W = e/sin\theta_W$.

It is instructive to work out the magnetic flux for the $SU(2)$ fields.
{}From (2.7) with (2.11), the net non-Abelian flux is:
$$
F_n = F_n |_S + F_n |_\Sigma = {{4\pi} \over {g}}
\eqno (2.12)
$$
just as we would expect for an ordinary $SU(2)$ monopole. That is,
the Dirac quantization condition works perfectly well for the $SU(2)$
field and the monopole charge is quantized in units of $4\pi /g$.
Another way of looking at (2.12) is to say that the electroweak
monopole is a genuine $SU(2)$ monopole in which there is a net emanating
$SU(2)$ flux. The structure of the theory, however, only permits a linear
combination of this flux and hypercharge flux to be long range and so
there is a string attached to the monopole. But this string is not
a Dirac string that surreptitiously returns the monopole flux; at best
it is a Dirac string only for the hypercharge part of the monopole field.
And so the electroweak monopole is a genuine topological monopole
in the $SU(2)$ sector of the model.
In particular, the magnetic charge on the monopole is conserved and
electroweak monopoles can only disappear by annihilating with
antimonopoles.

\noindent{\bf 3. Electroweak Dyons}

We now show that the electroweak model also admits dyon configurations.
To this end, we will write down dyonic configurations that
solve the asymptotic field equations.
Our analysis is similar to that in Ref. \cite{alford}, the only
difference being that whereas Alford et. al. linearized
their equations, we can work with the full non-linear equations. It should
also be mentioned that Nambu recognized the possibility of generalizing
the monopole solution with what he called ``external'' potentials.
Essentially, the dyon solution is an electroweak monopole together
with a particular external potential.

The classical field equations of motion for the bosonic sector of the
standard model of the electroweak interactions are:
$$
D^\mu D_\mu \Phi + 2\lambda \left ( \Phi^{\dag} \Phi -
                                 {{\eta^2} \over 2} \right ) \Phi = 0
\eqno (3.1)
$$
$$
D_\nu W^{\mu \nu a} = j^{\mu a} = {i \over 2} g \left [
   \Phi^{\dag} \tau^a D^\mu \Phi - (D^\mu \Phi )^{\dag} \tau^a \Phi
                                              \right ]
\eqno (3.2)
$$
$$
\partial_\nu Y^{\mu \nu} = j^{\mu Y} = {i \over 2} g' \left [
      \Phi^{\dag} D^\mu \Phi - (D^\mu \Phi )^{\dag} \Phi
                                                     \right ]
\eqno (3.3)
$$
where,
$$
D_\mu \Phi = \left [ \partial_\mu - {i \over 2} g W_\mu ^a \tau^a
                       - {i \over 2} g' Y_\mu \right ] \Phi
\eqno (3.4)
$$
and,
$$
D_\nu W^{\mu \nu a} =  \partial_\nu W^{\mu \nu a} +
                           g \epsilon^{abc} W_\nu ^b W^{\mu \nu c} \ .
\eqno (3.5)
$$

Denoting Nambu's monopole-string configuration by $({\bar \Phi},
{\bar W}_\mu ^a, {\bar Y}_\mu )$, the explicit fields for the monopole
in the asymptotic region are:
$$
{\bar \Phi} = {{\eta} \over {\sqrt{2}}}
          \pmatrix{ cos(\theta /2) \cr sin(\theta /2) e^{i\phi }} \
\eqno (3.6a)
$$
where, $\theta$ and $\phi$ are spherical coordinates centered on the
monopole. The gauge field configuration in the asymptotic region is:
$$
g {\bar W}_\mu ^a = - \epsilon^{abc} n^b \partial_\mu n^c + i cos^2 \theta_w
   n^a (\Phi^{\dag} ~ \partial_\mu \Phi - \partial_\mu \Phi^{\dag} ~\Phi )
\eqno (3.6b)
$$
$$
g' {\bar Y}_\mu = - i sin^2 \theta_w
 (\Phi^{\dag} ~ \partial_\mu \Phi - \partial_\mu \Phi^{\dag} ~\Phi ) \
\eqno (3.6c)
$$
where, $n^a$ is given in (2.2) and has the two very useful properties:
$$
n^a n^a = 1 \ , \ \ \ (n^a \tau^a + {\bf 1} ) \Phi = 2 {\bf Q} \Phi = 0
\eqno (3.7)
$$
where, ${\bf 1}$ is the $2\times 2$ unit matrix and ${\bf Q}$ is the
generator for the electromagnetic gauge transformations and is defined
to annihilate $\Phi$.

In writing (3.6), there is a subtlety which we should point out. One is
used to thinking of a monopole where the non-Abelian nature of the
fields is important only in a small core region and the non-Abelian
excitations fall off exponentially fast outside this core. Hence,
the asymptotic region in this case would be the region outside the
core. Here, however, there is a string sticking out of the monopole
and there is always a fraction of the region far from the monopole
in which $\Phi$ does not lie on the vacuum manifold. As we go
further away from the monopole, this fractional volume decreases
and our asymptotic approximations become valid over a larger fraction
of the asymptotic sphere. But the region where our approximations
are invalid only diminishes as a power law and not exponentially.
However, since we have introduced the denominator in (2.2),
$n^a$ has unit magnitude everywhere - even inside the string. In
addition, the derivatives of $n^a$ inside the string fall off as
$1/R$ where $R$ is the distance from the monopole and $n^a$ can
be taken to be constant inside the string in the asymptotic region.

Now we make an ansatz that will describe an electroweak dyon connected by
a semi-infinite $Z$ string. This is:
$$
\Phi = {\bar \Phi}
\eqno (3.8)
$$
$$
W_\mu ^a = {\bar W}_\mu ^a - \delta_\mu ^t {{ n^a {\dot \gamma}} \over
                                              cos\theta_W}
\eqno (3.9a)
$$
$$
Y_\mu = {\bar Y}_\mu - \delta_\mu ^t {{\dot \gamma} \over
                                              sin\theta_W}
\eqno (3.9b)
$$
where, $\gamma = \gamma (\vec x ,t)$ and overdots denote partial
time derivatives.

We can work out the field strengths for the dyon-string ansatz. The
change in the field strengths are:
$$
\delta W_{ij}^a = 0 = \delta Y_{ij}
\eqno (3.10)
$$
$$
\delta W_{it}^a = - { {{\bar D}_i (n^a {\dot \gamma} )} \over
                      {cos\theta_W}}
\ , \ \ \
\delta Y_{it} = - {{\partial_i {\dot \gamma}} \over {sin\theta_W}}
\eqno (3.11)
$$
where,
$$
{\bar D }_\mu (n^a {\dot \gamma } ) \equiv \partial_\mu (n^a {\dot \gamma} )
           + g \epsilon^{abc} {\bar W}_\mu ^b ( n^c {\dot \gamma} )
\eqno (3.12)
$$
is the covariant derivative of $n^a {\dot \gamma}$ with the monopole-string
gauge field of eqn. (3.6).

Now we first check if the $\Phi$ eqn. (3.1) is satisfied by the dyon-string
ansatz in the asymptotic region. For this, we use (3.7) and find that
$$
D_\mu \Phi = {\bar D}_\mu {\bar \Phi}
\eqno (3.13)
$$
where, ${\bar D}_\mu$ is defined as in (3.4) but with barred gauge
fields. This leads to
$$
D^\mu D_\mu \Phi = {\bar D}^\mu {\bar D}_\mu {\bar \Phi}
\eqno (3.14)
$$
upon using ${\bar D}_t {\bar \Phi} = 0$. Then it trivially follows that the
$\Phi$ equation of motion with the dyon-string ansatz is satisfied, if it is
satisfied by the monopole-string ansatz.

Next, we work out the change in the currents due to the extra terms in
(3.9). A little algebra yields
$$
\delta j^{\mu a} = 0 = \delta j^{\mu Y} \ .
\eqno (3.15)
$$
Inserting these in (3.2) and (3.3) we find that the $W$ gauge field
equation leads to:
$$
{\bar D}^i {\bar D}_i (n^a \dot \gamma ) = 0
\eqno (3.16)
$$
$$
{\partial_t} {\bar D}^i (n^a \dot \gamma ) = 0
\eqno (3.17)
$$
while the $Y$ equation gives:
$$
\partial^i \partial_i \dot \gamma = 0
\eqno (3.18)
$$
$$
\partial_t \partial^i \dot \gamma = 0 \ .
\eqno (3.19)
$$

The first two equations ((3.16) and (3.17)) of these four equations for
$\gamma$, are consistent with the second two when we use the explicit
expressions in (3.6) for the monopole-string configuration.
To see this, we use (3.6) and obtain
$$
{\bar D}_i (n^a \dot \gamma ) = n^a {\partial_i} \dot \gamma \ .
\eqno (3.20)
$$
With this relation, (3.16) and (3.17) reduce to (3.18) and (3.19).

In obtaining (3.20) we have made use of the explicit form of
the monopole-string field configuration (eqn. (3.6)) for the first time.
and the question arises if (3.20) holds within the string too where (3.6)
does not apply. This is indeed the case because $n^a$ is constant and
${\bar W}_\mu ^a$ is parallel to $n^a$ inside the string in the asymptotic
region (see the discussion following (3.7)). Using these facts, $n^a$ pulls
out of the partial derivative in ${\bar D}_i$ and the cross-product term in
the covariant derivative (eqn. (3.12)) does not contribute. So (3.20) is
valid even inside the string.

Next, we use separation of variables and write
$$
\gamma = \xi (t) f(\vec x ) \ .
\eqno (3.21)
$$
This leads to
$$
{\ddot \xi} = 0 \ , \ \ \ \nabla ^2 f = 0 \ ,
\eqno (3.22)
$$
with general solutions that can be found in many text-books\refto{jackson}.
The particular solution that we will be interested in is the
solution that gives a dyon. Hence, we take:
$$
\xi = \xi_0 t \ , \ \ \
 f(r) = - {{q sin\theta_W cos\theta_W} \over {4\pi \xi_0}} {1 \over r} \ ,
\eqno (3.23)
$$
where, $\xi_0$ is some constant. Now, using (2.3), together with (3.7),
(3.9) and (3.23) we then get the dyon electric field:
$$
{\vec E}_A = {q \over {4\pi}} {{\vec r} \over {r^3}} \ .
\eqno (3.24)
$$

For a long segment of string, the monopole and the antimonopole at
the ends are well separated and we can repeat the above analysis for both
of them independently. Therefore, the electric charge on the antimonopole
at one end of a $Z$ string segment is uncorrelated with the charge on the
monopole at the other end of the string. This means that we can have
dyons of arbitrary electric charge at either end of the string.

This completes our construction of the dyon-string system in the electroweak
model. As of now, the charge $q$ on the dyon is arbitrary; we will quantize
it in the next section.

\noindent{\bf 4. Charge Quantization}

Consider a segment of $Z$ string which has electroweak dyons at either
end with magnetic and electric charges
$$
(m, q_1) \ \ {\rm and} \ \ (-m, q_2) \ .
\eqno (4.1)
$$
The angular momentum of this system is given by the sum of the angular
momenta in the $SU(2)$ and hypercharge fields. However, according
to our discussion in Sec. 2, the monopole resides entirely in the $SU(2)$
sector and so there can be no angular momentum due to the hypercharge
field. Furthermore, the dyons have $SU(2)$ magnetic and $SU(2)$ electric
charges given by
$$
\left ({{4\pi} \over {g}} , {{q_1} \over {sin\theta_W}} \right )
\ \ {\rm and} \ \
\left ({{- 4\pi} \over {g}} , {{q_2} \over {sin\theta_W}} \right )\ .
\eqno (4.2)
$$
And the $SU(2)$ fields due to these charges are long range since the fields
occur as part of the electromagnetic flux from the monopole. So the
angular momentum in the field is given by the usual ``cross-product''
of the magnetic and electric charges:
$$
L = {{4\pi} \over g} {{q_2} \over {sin\theta_W}} -
     {{- 4\pi} \over g} {{q_2} \over {sin\theta_W}}
= {{ 4\pi (q_1 + q_2) } \over {g sin\theta_W}}
\eqno (4.3)
$$
Quantizing the angular momentum in units of $4\pi$ ($\hbar = 1$)
and using $e= g~sin\theta_W$ yields
$$
q_1 + q_2 = n e
\eqno (4.4)
$$
where, $n$ is an integer.

The charges $q_1$ and $q_2$ are independent of each other and if we
have CP invariance, we can consider a CP transformed dyon on one end
of the string. This will change $q_2$ to $- q_2$ and hence, the
charge on each of the dyons is quantized as $ne$ or $(n+1/2)e$ if our
theory is invariant under CP. If we further assume that the zero electric
charge dyon is in the spectrum, then the quantization is in units of $e$.

One way to have CP violation in the theory is to have a $\theta$ term
in the action. For the standard model of the electroweak interactions
(without fermions!) this would be
$$
S_\theta = {{g^2 \theta} \over {32 \pi^2}} \int d^4 x
            W_{\mu \nu}^a {\tilde W}_{\lambda \sigma}^a \
= {{g^2 \theta} \over {8 \pi^2}} \int d^4 x {\vec E}^a \cdot {\vec B}^a \ .
\eqno (4.5)
$$
Now, in the asymptotic region, only the $SU(2)$ fields have a
net flux from the monopole (see (2.12)). So we can write:
$$
\nabla \cdot {\vec B}_n = {{4\pi} \over g} \delta^3 (\vec x ) \ .
\eqno (4.6)
$$
where, the subscript $n$ denotes a combination of $SU(2)$ field strengths
parallel to $n^a$.
Following Coleman's derivation\refto{sc} of the Witten effect, an integration
by parts of (4.5) then shows that the dyon acquires an extra $SU(2)$
charge which is $g \theta /2\pi$. But this is exactly an electric
charge $e\theta /2\pi$ where $e = g~sin\theta_W$ (see (2.1)). Hence, the
electric charge on the dyon in the presence of a $\theta$ term is
$$
q = \left ( n + {\theta \over {2\pi}} \right ) e \ .
\eqno (4.7)
$$
This agrees fully with the standard result for dyons and it is reassuring
to see that the periodicity under $\theta \rightarrow \theta + 2\pi$
is intact.

It should be mentioned that, even though the electric charge on an electroweak
dyon can be fractional as in (4.7), the total electric charge on the
dyon-string system is always integral because the CP violating fractional
charge on the monopole is equal and opposite to that on the antimonopole.

\noindent{\bf 5. Discussion}

The standard model contains fermions in addition to the bosonic sector
that we have considered so far. With the inclusion of fermions it is
known that the $\theta$ term can be gauged away and so the Witten
charge must disappear. However, it
is also known that the standard model contains CP violation in the
fermionic sector via the Kobayashi-Maskawa matrix. As suggested by
Witten, any CP violation in a model with monopoles will feed into the dyon
charge spectrum. Hence, electroweak monopoles should feel the KM matrix
and obtain a fractional charge. While the existence of such a fractional
charge seems believable, it is a much harder problem to actually compute
what the value of the charge is. This burning question is currently being
investigated.

The presence of charge on electroweak monopoles is likely to have
a CP violating influence on the evolution of electroweak strings.
In connection with our earlier finding that electroweak strings
can carry baryon number\refto{tvgf}, the CP violating dynamics of such
strings could lead to the potentially important phenomenon that the decay of
baryon number carrying strings preferentially produces baryons rather
than antibaryons.

\

\

\noindent {\it{Acknowledgements:}}

I am grateful to Sidney Coleman for several insights. This work was
supported by the NSF.

\vfill

\references

\refis{ew} E. Witten, Nucl. Phys. B 249, 557 (1985).

\refis{tv1}  T. Vachaspati, Phys. Rev. Lett. 68, 1977 (1992);
69, 216(E) (1992); Nucl. Phys. B397, 648 (1993).

\refis{sc} S. Coleman, ``The Magnetic Monopole Fifty Years Later'',
Erice Lectures (1981).

\refis{alford} M. Alford et. al., Nucl. Phys. B 384, 251 (1992).

\refis{jackson} For example, ``Classical Electrodynamics'', J. D. Jackson,
Wiley (1975).

\refis{yn} Y. Nambu, Nucl. Phys. B130, 505 (1977).

\refis{hnpo} H. B. Nielsen and P. Olesen, Nucl. Phys. B 61, 45 (1973).

\refis{tvmb} T. Vachaspati and M. Barriola, Phys. Rev. Lett. 69,
1867 (1992).

\refis{tvgf} T. Vachaspati and G. B. Field, ``Electroweak String
Configurations with Baryon Number'', TUTP-94-1 (1994).

\endreferences

\vfill
\eject


\endjnl
\end